# MUON COLLIDERS AND NEUTRINO FACTORIES*

S. Geer, Fermilab, P.O. Box 500, Batavia, IL 60510, U.S.A.

*Abstract*

Over the last decade there has been significant progress in developing the concepts and technologies needed to produce, capture and accelerate $O(10^{21})$ muons/year. This development prepares the way for a new type of neutrino source (Neutrino Factory) and a new type of very high energy lepton-antilepton collider (Muon Collider). This article reviews the motivation, design and R&D for Neutrino Factories and Muon Colliders.

## INTRODUCTION

The muon, which can be thought of as a heavy electron, lives just long enough ($\tau_0 = 2\mu s$) to enable it to be accelerated to high energy before it decays into an electron, a muon-type neutrino and an electron-type antineutrino. Over the last decade there has been significant progress in developing the concepts and technologies needed to produce, capture and accelerate $O(10^{21})$ muons/year. This prepares the way for (i) a Muon Collider (MC) [1] in which $\mu^+$ and $\mu^-$ are brought to collision in a storage ring, and (ii) a Neutrino Factory (NF) [2] in which high energy muons decay within the straight sections of a storage ring to produce a beam of neutrinos and anti-neutrinos.

## MUON COLLIDER MOTIVATION

Over the years $e^+e^-$ colliders, have played an important role in establishing and testing the Standard Model. The physics program that could be pursued by a new lepton collider ($e^+e^-$ or $\mu^+\mu^-$) has captured the imagination of the high energy physics community. With sufficient energy and luminosity this new accelerator would facilitate:

- understanding the mechanism behind mass generation and electroweak symmetry breaking;
- searching for, and perhaps discovering, super-symmetric particles and confirming their nature;
- hunting for extra space-time.

Within a few years results obtained from the Large Hardron Collider at CERN are expected to more precisely establish the desired lepton collider energy, and whether the physics program can be begun with a lower energy ($\sqrt{s} \sim 0.5$ TeV) collider, or whether we must go straight to multi-TeV energies to make contact with the physics. In either case, it is likely that multi-TeV lepton colliders will eventually be needed.

Both $e^+e^-$ and $\mu^+\mu^-$ colliders have been proposed as possible candidates for a multi-TeV lepton collider. However, a relativistic particle undergoing centripetal acceleration radiates at a rate proportional to the fourth power of the Lorentz factor ($\gamma^4$). This poses a challenge for multi-TeV $e^+e^-$ colliders, which cannot be circular, but must have a linear geometry and, with practical acceleration schemes, be tens of km long. Furthermore, beam-beam effects at the collision point induce the electrons and positrons to radiate, which broadens the colliding beam energy distributions. Since $(m_\mu/m_e)^4 = (207)^4 = 2\times 10^9$, all of these radiation-related effects can be mitigated by using muons instead of electrons. A multi-TeV $\mu^+\mu^-$ collider can be circular and therefore have a compact geometry that will fit on existing accelerator sites. Furthermore, since TeV scale muons can be bent in arcs, muilti-pass acceleration can be used, promising a cost-effective way of accelerating to TeV energies.

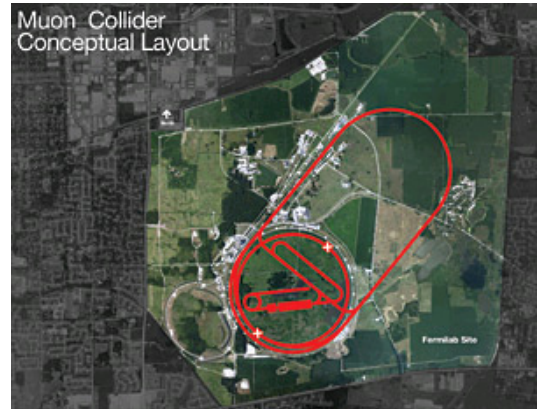

Figure 1: Footprint of a 3 TeV Muon Collider complex on the 6,800 acre Fermilab site.

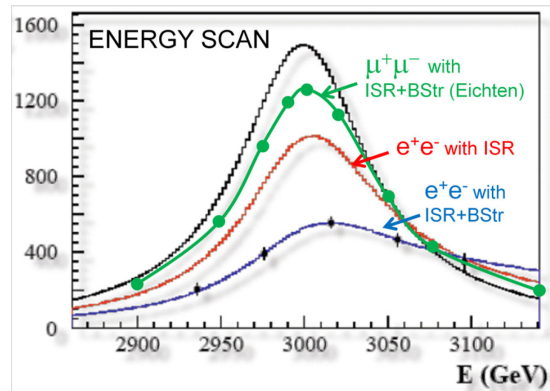

Figure 2: Comparison of collision energy spreads for 3 TeV $e^+e^-$ and $\mu^+\mu^-$ colliders. Top black curve is before radiative effects are included. Green curve shows muon collider energy spread including iniial state radiation (ISR) and beamsstrahlung (BS), Red curve shows the

---

* Work supported at the Fermi National Accelerator Laboratory, which is operated by the Fermi Research Association, under contract No. DE-AC02-76CH03000 with the U.S. Department of Energy.

impact of ISR on an e⁺e⁻ collider, and the bottom blue curve shows the combined effects of ISR+BS on an e⁺e⁻

## THE CHALLENGE

There are some significant challenges that must be collider.

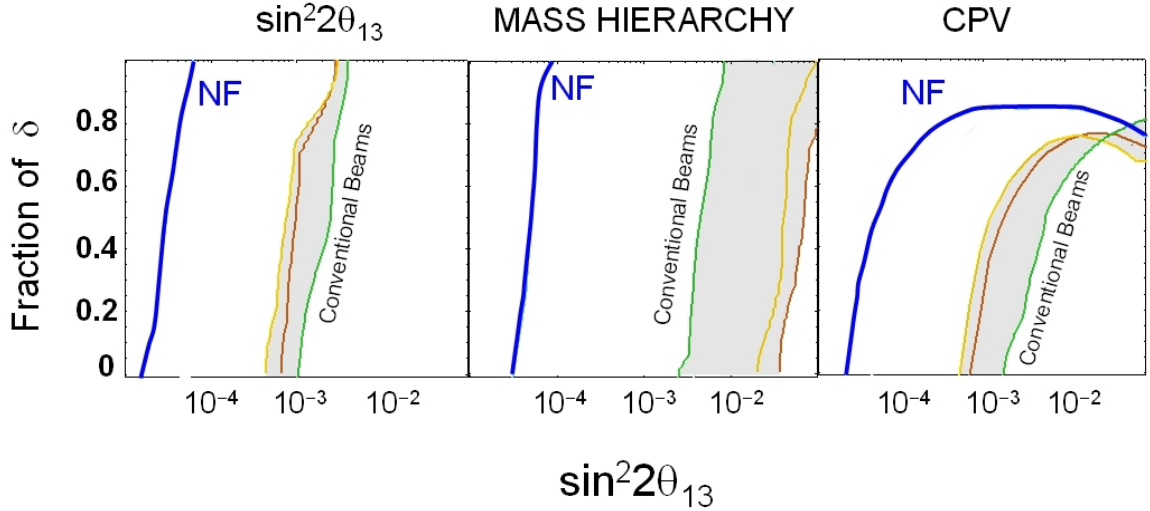

Figure 3: Results from the [4]. As a function of $\sin^2 2\theta_{13}$, the fraction of all possible values of $\delta$ for which a discovery could be made at the 3σ level or better at a 25 GeV NF (blue curves) and for a selection of possible future conventional beam experiments (curves in gray bands).

The expected footprint for a 3 TeV muon collider is shown in Fig. 1, and beam energy spreads for multi-TeV e⁺e⁻ and μ⁺μ⁻ colliders are compared in Fig. 2.

## NEUTRINO FACTORY MOTIVATION

A Neutrino Factory would produce a unique neutrino beam generated by the decays of a well formed beam of mono-energetic muons. This is very different from a conventional neutrino beam formed by the decays of a broad spectrum of charged pions in a large aperture decay channel. The NF offers a beam with well known flux and spectrum, and with electron-type, as well as muon-type neutrinos and anti-neutrinos. The resulting physics reach has been well studied [4]. If the unknown neutrino mixing angle $\theta_{13}$ is very small, a 25 GeV NF would offer exquisite sensitivity for discovering a non-zero $\theta_{13}$, establishing the pattern of neutrino mass (the mass hierarchy), and searching for leptonic CP violation. If $\theta_{13}$ is "large", a ~5 GeV NF would offer outstanding precision for following-up the initial discoveries and searching for something new.

overcome before a NF and/or a MC can be built. The challenges arise because:

- Muons are produced in charged pion decays, and are therefore born within a large 6D phase-space. For a NF, the transverse phase space must be cooled by a factor of a few in each transverse plane, to fit within the acceptance of an accelerator. For a MC, the 6D phase-space must be cooled by $O(10^6)$ to obtain sufficient luminosity.
- Muons decay. Everything must be done fast. In particular, a new beam cooling technique (ionization cooling) is required. We must also deal with the decay electrons, and for a MC with an energy of 3 TeV or greater, attention must be paid to the decay neutrinos.
- We need a lot of muons. In practice, this means we must start with a MW-scale proton source, for example: a 4 MW beam at 8 GeV.

## DESIGN

Muon Collider and Neutrino Factory accelerator complexes are shown schematically in Fig. 4. At the front-end, both the NF and MC require similar, perhaps identical, intense muon sources, and hence there is significant overlap in NF and MC R&D. The muon source is designed to deliver $O(10^{21})$ low energy muons per year within the acceptance of an accelerator, and consists of (*i*) a multi-MW proton source delivering a multi-GeV proton beam onto a pion production target, (*ii*) a high-field target solenoid that radially confines the secondary charged pions, (*iii*) a long solenoidal channel in which the pions decay to produce positive and negative muons, (*iv*) a system of rf cavities that capture the muons in bunches and reduce their energy spread (phase rotation), and (*v*) a muon ionization cooling channel that reduces the transverse phase space occupied by the beam by a factor of a few in each transverse direction. At this point the beam will fit within the acceptance of an accelerator for a NF. However, to obtain sufficient luminosity, a MC requires further muon cooling. In particular, the 6D phase-space must be reduced by $O(10^6)$, which requires a longer and more ambitious cooling channel. Finally, in both NF and MC schemes, after the cooling channel the muons are accelerated to the desired energy and injected into a storage ring. In a NF the ring has long straight sections in which the neutrino beam is formed by the decaying muons. In a MC, positive and negative muons are injected in opposite directions and collide for about 1000 turns before the muons decay.

## (a) Neutrino Factory

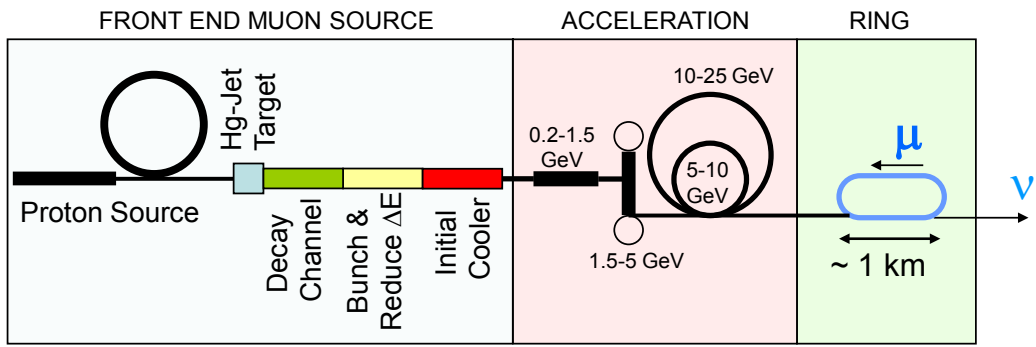

## (b) Muon Collider

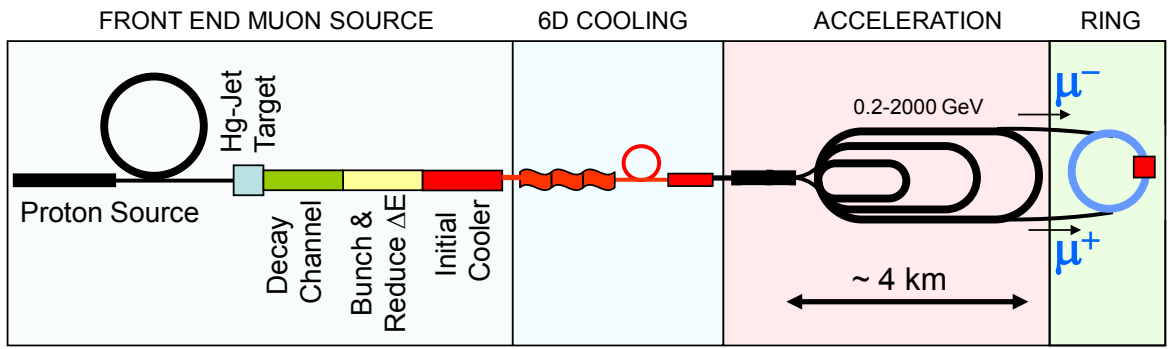

Figure 4: Neutrino Factory [2,3] and Muon Collider [1] schematics.

## FRONT END

At the end of the decay channel the daughter muons have drifted some tens of meters, resulting in a time-energy correlation with the high-energy particles leading the low-energy particles. The decay channel is followed by a buncher section that uses rf cavities to form the beam into a bunch train, and a phase-energy rotating section that decelerates the early-rf-phase high energy bunches and accelerates the late-rf-phase low energy bunches, so that each bunch has the same mean energy. Present designs deliver a bunch train that is 50m long, captured within a 2T solenoid channel.

The buncher parameters are determined by considering reference particles (1, 2) with velocities β₁ and β₂. The rf voltages are increased along the channel, with frequencies $f_{rf}$ and phases set to place 1 and 2 at the center of bunches. This can be accomplished if the rf wavelength $\lambda_{rf}$ increases along the buncher [5]:

$$N_B \lambda_{rf}(s) = N_B \frac{c}{f_{rf}(s)} = s\left(\frac{1}{\beta_2} - \frac{1}{\beta_1}\right)$$

where s is the total distance from the target and $N_B$ is an integer. In the present design, at the end of the channel all bunches have a mean momentum ~230 MeV/c, with μ⁺ and μ⁻ bunches interleaved within the rf cycle. The reduction in the overall energy spread effectively increases the number of useful muons by a factor of ~4.

The number of muons accepted by the downstream accelerators can be further increased by reducing the emittance in each transverse direction by a factor of a few. This can be accomplished using "ionization cooling" in which the muons lose energy by ionization as they pass through an absorber. This reduces their momenta in the longitudinal- and transverse-directions. An rf cavity then replaces the lost energy by reaccelerating in the longitudinal direction. After repeating the process many times, the transverse momenta (and transverse emittances) are reduced. The rate at which the normalized transverse emittance $\varepsilon_{xN} = \beta\gamma\, \varepsilon_x$ changes as muons with energy $E_\mu$ (GeV) lose energy by ionization loss $dE_\mu/ds$ within material with radiation length $L_R$ is given by:

$$\frac{d\varepsilon_{Nx}}{ds} = -\frac{dE_\mu}{ds}\frac{\varepsilon_{Nx}}{E_\mu} + \frac{\beta_\perp (0.014)^2}{2 E_\mu m_\mu L_R},$$

where $\beta_\perp$ characterizes the focusing strength at the absorber. The second term describes heating due to scattering in the absorber which ultimately limits the cooling process. To minimize the impact of scattering it is desirable to use a low Z (high $L_R$) absorber (e.g. liquid hydrogen or LiH) and to focus the muons strongly (small $\beta_\perp$) so that the focusing angles are much larger than typical scattering angles. The present baseline cooling channel design consists of a sequence of LiH absorbers and 201 MHz rf cavities within a lattice of solenoids that provide the required focusing. Simulations show that the cooling channel increases the number of useful muons by a factor of ~2.

To provide a proof-of-principle demonstration, the international Muon Ionization Cooling Experiment (MICE [6]) at RAL is preparing to test an ionization cooling channel cell in a muon beam. MICE will measure the response of individual muons to the cell as a function of the incident muon parameters (momentum, position, direction) and the various channel parameters (absorber type, magnetic fields, rf parameters). The initial phase of the experiment, which establishes the muon beam and measurement systems, has begun. It is anticipated that MICE will be completed by ~2013.

The beam exiting the front end has $\varepsilon_{\perp N}$=6mm and $\varepsilon_{\parallel N}$=25mm, and will fit within the acceptance of an accelerator. It is therefore suitable for acceleration to a few GeV or a few tens of GeV for a NC. However, a more aggressive 6D cooling channel is required to meet the O($10^{34}$) cm⁻²s⁻¹ luminosity goal for a TeV-scale MC.

## RF IN MAGNETIC FIELDS

The bunching, phase rotation, and cooling channel designs require high gradient normal conducting rf cavities operating in a magnetic channel. The preferred design exploits the penetrating nature of muons by using cavities in which the normally open rf cells are closed with thin conducting windows. At fixed peak power this doubles the effective accelerating gradient, and hence halves the required number of rf power sources. Thin beryllium windows for this purpose have been demonstrated in an 805 MHz test cavity. However, tests have shown that when this type of cavity is operated within a multi-Tesla co-axial solenoid the maximum rf gradient that can be achieved before breakdown is significantly reduced. It is possible that, with further R&D, surface treatments can be found to mitigate this effect. However, other solutions have also been proposed including:

(i) Using cavities filled with high pressure hydrogen gas. An 805 MHz cell has been built and tested in a high field solenoid. No appreciable degradation of performance was observed with increasing magnetic field. In the coming months this technology will be tested in the presence of an intensely ionizing beam. It is possible that the ionization created in the cavity will limit its performance.

(ii) Using "magnetically insulated" cavities. The magnetic field is designed so that it is parallel to surfaces where the rf gradients are maximum. This is expected to prevent energetic electrons from hitting these surfaces and causing problems.

(iii) Designing cooling channels in which the cavities are in regions of low magnetic field. This is not a preferred solution since it will mean longer less efficient channels.

Within the next couple of years the ongoing R&D is expected to determine which of these options are viable.

## 6D COOLING

The beam exiting the front end will fit within the acceptance of an accelerator. However, far more cooling is required to meet the luminosity requirements for a MC. Since ionization cooling works only in the transverse plane, to cool in 6D requires mixing of the degrees of freedom in the cooling channel. One way to do this is to employ a solenoid channel in which the coils (and beam) have a helical geometry, with carefully chosen helix parameters. This has been simulated and appears to work well provided rf and magnet parameters can be achieved.

## ACCELERATION

Various acceleration schemes have been studied for a 25 GeV NF. Typically they begin with a linear "pre-accelerator" that accelerates the beam to about 1 GeV. The muons are then sufficiently relativistic to use a Recirculating Linear Accelerator (RLA) in which arc-sections return the muons to the same linac several times. Higher energies can be obtained using further RLAs and/or FFAG (Fixed Field Alternating Gradient) accelerators. The present baseline scheme uses a "non-scaling FFAG" to raise the energy to 25 GeV. The EMMA experiment [7] at Daresbury has been designed to study non-scaling FFAG beam dynamics, which are interesting because the particles are accelerated out of the rf bucket. EMMA results will enable the attractiveness of this particular scheme to be better assessed.

The MC could use the same initial acceleration scheme as the NF. To get from 25 GeV to TeV energies, a sequence of RLAs might then be used. However, rapid cycling synchrotrons are likely to be more cost effective, and R&D towards developing the required magnets is ongoing.

## MAP

The Neutrino Factory and Muon Collider Collaboration (NFMCC) has been pursuing muon accelerator R&D since 1996. The initial work on the overall Muon Collider concept resulted in the "Muon Collider Feasibility Study Report" in June 1996 [1]. The Neutrino Factory concept emerged in 1997 [2]. Since 1997 the NFMCC has pursued both NF and MC design and simulation studies, together with component development and proof-of-principle demonstration experiments. In late 2006, the Muon Collider R&D effort was complemented by the addition of the Muon Collider Task Force (MCTF) centered at Fermilab. This doubled the support in the U.S. for NF and MC R&D. By 2009 the NFMCC + MCTF community, together with their international partners (MICE, MERIT [8], IDS-NF [9]) had made significant progress, completing a series of NF design feasibility studies [3] and the proof-of-principle target experiment MERIT, launching MICE and a hardware component development program, building the Mucool Test Area at Fermilab, and making progress with 6D cooling channel studies.

Given these achievements, in October 2009 the DOE requested the Fermilab Director to put in place and host a new national Muon Accelerator R&D organization (Muon Accelerator Program, MAP [10]) to replace and streamline the NFMCC+MCTF activities, with an expectation of increased funding. MAP is now in place and functioning. A MAP R&D plan has been submitted and reviewed. The main R&D deliverables of MAP over the next few years will be:

1. A Design Feasibility Study Report (DFSR) for a multi-TeV MC including an end-to-end simulation of the MC accelerator complex using demonstrated, or likely soon-to-be-demonstrated, technologies, an indicative cost range, and an identification of further technology R&D that should be pursued to improve the performance and/or the cost effectiveness of the design.
2. Technology development and system tests that are needed to inform the MC-DFSR studies, and enable an initial down-selection of candidate technologies for the required ionization cooling and acceleration systems.
3. Contributions to the International Neutrino Factory Design Study (IDS-NF) to produce a Reference Design Report (RDR) for a NF by 2013. The emphasis of the proposed U.S. participation is on: *a*) design, simulation and cost estimates for those parts of the NF front-end that are (or could be) in common with a MC; *b*) studying how the evolving Fermilab proton source can be used for the Neutrino Factory RDR design; and *c*) studying how the resulting NF would fit on the Fermilab site.

## SUMMARY

MC and NF R&D is well motivated, and is receiving increased support and attention. Many of the subsystems for these new types of facility require "linac technologies with a twist" (literally in the case of a 6D cooling channel). In particular, (i) a muon cooling channel can be thought of as a linac filled with material, (ii) the linac structures must cope with the decay electrons, and (iii) high gradient NCRF is desired operating in magnetic fields of a few Tesla. In the next few years we can anticipate the IDS-NF community delivering a NF RDR, and the MAP community delivering a MC design feasibility study report. Armed with these design and feasibility studies, and results from the hardware R&D that is being pursued in parallel, we can hope that these new facilities will become options for particle physics.